\documentclass{amsart}

 \theoremstyle{definition}
 
 \theoremstyle{remark}
 
\numberwithin{equation}{section}

 \newcommand{\h}{\mathfrak{H}}

\usepackage{amsfonts}
\usepackage{amssymb}
\usepackage{graphicx, color}
\usepackage{hyperref}

\begin{document}


\title[Occupation Probabilities]{On the identification of the ground state
based on occupation probabilities:\\An investigation of Smith's
apparent counterexamples~\footnote{This is an edited version, with
updated references, of an earlier reply to Smith in July, 2005.}}


\author{Tien D. Kieu}
\email{kieu@swin.edu.au}
\address{Centre for Atom Optics and Ultrafast Spectroscopy, and ARC
Centre of Excellence for Quantum-Atom Optics, Swinburne University
of Technology, Hawthorn 3122, Australia}


\begin{abstract}
We study a set of truncated matrices, given by
Smith~\cite{Smith2005}, in connection to an identification criterion
for the ground state in our proposed quantum adiabatic algorithm for
Hilbert's tenth problem. We identify the origin of the trouble for
this truncated example and show that for a suitable choice of some
parameter it can always be removed.	 We also argue that it is only
an artefact of the truncation of the underlying Hilbert spaces,
through showing its sensitivity to different boundary conditions
available for such a truncation. It is maintained that the
criterion, in general, should be applicable provided certain
conditions are satisfied.  We also point out that, apart from this
one, other criteria serving the same identification purpose may also
be available.
\end{abstract}

\maketitle

In a proposal of a quantum adiabatic algorithm for Hilbert's tenth
problem~\cite{thisissue}, we employ an adiabatic process with a
time-dependent Hamiltonian
\begin{eqnarray}
{\h}(t) &=& (1-t/T) H_I + (t/T) H_P. \label{Hamiltonian}
\end{eqnarray}
Here $t$ is time and this Hamiltonian metamorphoses from $H_I$ when
$t=0$ to $H_P$ when $t=T$.  The final Hamiltonian $H_P$ encodes the
Diophantine equation in consideration, while the initial $H_I$ is
universal and independent of the Diophantine equation, except only
on its number of variables $K$.  The process is captured by the
Schr\"odinger equation
\begin{eqnarray}
\partial_t |\psi(t)\rangle &=& -i\h(t)|\psi(t)\rangle,\\
|\psi(0)\rangle &=& |\{\alpha\}\rangle,\nonumber
\label{Schroedinger}
\end{eqnarray}
where $|\{\alpha\}\rangle$ is the ground state of the initial
infinite-dimensional Hamiltonian $H_I$, which we choose to be
\begin{eqnarray}
H_I &=& \sum_{i=1}^K (a^\dagger_i - \alpha_i^*)(a_i - \alpha_i).
\label{HI}
\end{eqnarray}
This choice is not unique, but with it, $|\{\alpha\}\rangle$ is then
the Cartesian product, of $K$ factors, of the well-known coherent
states in quantum physics.

In order to identify the ground state at the final time $t=T$, we
have shown in the case of 2-dimensional spaces~\cite{kieuFull} that
if we could choose $\alpha$ such that $\left|\langle
\psi(0)|\{n\}\rangle\right|^2<0.5$, for all $|\{n\}\rangle$ which
are the Fock states and also the eigenstates of $H_P$, then
\begin{eqnarray}
{\mbox {\rm If }}\left|\langle \psi(T)|\{n^{(0)}\}\rangle\right|^2 >
\frac{1}{2} &\Rightarrow& |\{n^{(0)}\}\rangle \; {\mbox  {\rm  is
the ground state of }} H_P, \label{criterion}
\end{eqnarray}
assuming such a ground state is non-degenerate, and provided an
extra condition has to be satisfied, for all $0<t<T$, namely
\begin{eqnarray}
\langle e(t) | H_P - H_I | g(t)\rangle &\not=& 0, \label{condition}
\end{eqnarray}
where $|g(t)\rangle$ and $|e(t)\rangle$ are, respectively, the
instantaneous ground state and the first excited state of $\h(t)$ at
the time $t$. The {\em violation} of that condition at some $t_0$ is
equivalent to
\begin{eqnarray}
{\mbox {\rm Both }}&&\langle e(t_0) | H_P | g(t_0)\rangle = 0,
\label{condition1}\\
{\mbox {\rm And }}&&\langle e(t_0) | H_I | g(t_0)\rangle = 0.
\label{condition2}
\end{eqnarray}

In two dimensions this criterion for the ground-state identification
can easily and always be satisfied for $H_I$ and $H_P$
noncommutting, which is automatic for our choice of $H_I$.	For
higher dimensions, the non-commutativity of the Hamiltonians is no
longer a sufficient condition for~(\ref{condition}).	And we need
make sure that the condition~(\ref{condition}) is observed.

Recently, Smith has claimed to have three
counterexamples~\cite{Smith2005} against the
criterion~(\ref{criterion}) above.	Of the three provided, actually
there is only one that is relevant and fits the form of our
time-dependent Hamiltonian~(\ref{Hamiltonian}) in the case of one
variable, {\em but truncated to five dimensions} and with the
diagonal term $(1-|\alpha|^2){\bf 1}$ added to~(\ref{HI}),
\begin{eqnarray}
H_I^S= \left(
\begin{array}{ccccc}
1& -\alpha^*& 0& 0& 0\\
-\alpha& 2& -\alpha^*\sqrt{2}& 0& 0\\
0& -\alpha\sqrt{2}& 3& -\alpha^*\sqrt{3}& 0\\
0& 0& -\alpha\sqrt{3}& 4& -2\alpha^*\\
0& 0& 0& -2\alpha& 5
\end{array}
\right);&& H_P^S= \left(
\begin{array}{ccccc}
2& 0& 0& 0& 0\\
0& 4& 0& 0& 0\\
0& 0& 5& 0& 0\\
0& 0& 0& 3& 0\\
0& 0& 0& 0& 1
\end{array}
\right) .\label{matrices}
\end{eqnarray}
Smith has chosen the particular values $\alpha=1$ and $T = 13.3444$.
In this note we shall only investigate this example; other
criticisms by Smith in~\cite{Smith2005} also overlap with others',
our replies to which have been given in~\cite{thisissue}.

\begin{figure}
\begin{center}
\includegraphics{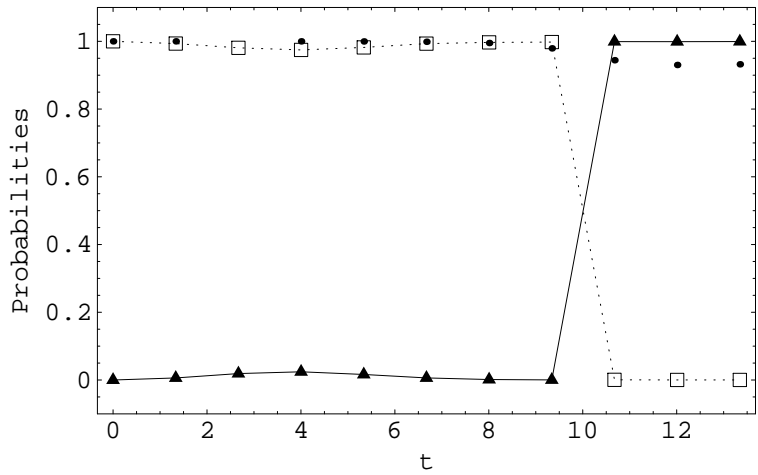}
\caption{\label{Fig2}The occupation probabilities of the
instantaneous ground state (unfilled boxes) and first excited state
(triangles) as functions of time, for $\alpha=1$. The excited-state
probability rises suddenly from near zero to close to one around
$t\approx 10$, which corresponds to the near miss of these two
energy levels as shown in Fig.~\ref{Fig3}.  The filled circles are
for the instantaneous ground-state occupation for $\alpha = 3$,
which shows no sudden change. See text for further explanation.}
\end{center}
\end{figure}

Indeed, when we solve the Schr\"odinger
equation~(\ref{Schroedinger}), starting with the exact ground state
of $H_I^S$, we obtain at time $t=0$ no Fock-state occupation
probabilities is more than 0.5, but at $t=T$, the {\em first
excited} Fock state has a probability of 0.999323 -- apparently
contradicting the criterion~(\ref{criterion}). We have depicted the
occupation probabilities of the {\em instantaneous} ground and first
excited states as functions of time in Fig.~\ref{Fig2}.  At first,
the ground-state occupation is close to one, as it should be because
we start the system in the initial ground state.  Then, suddenly at
around $t\approx 10$ this probability plummets to zero, accompanying
by a stellar rise of that of the first excited state.
%

\begin{figure}
\begin{center}
\includegraphics{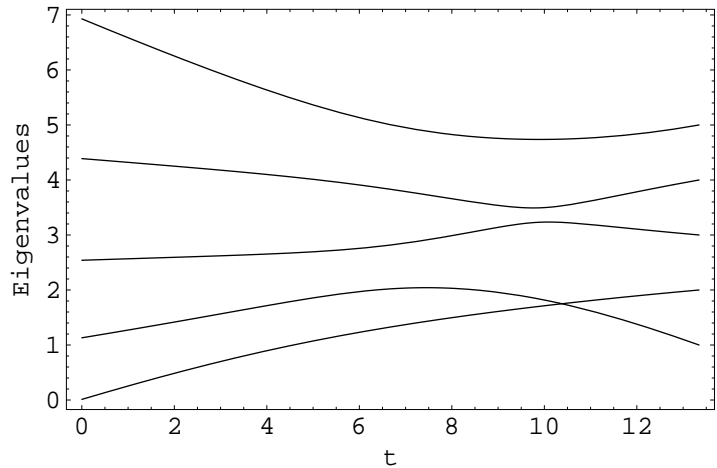}
\caption{\label{Fig3}The spectral flow for $\alpha =1$.  The
apparent crossing of the lowest two energy levels just after
$t\approx 10$ is actually an avoided crossing as shown in
Fig.~\ref{Fig4}.}
\end{center}
\end{figure}

\begin{figure}
\begin{center}
\includegraphics{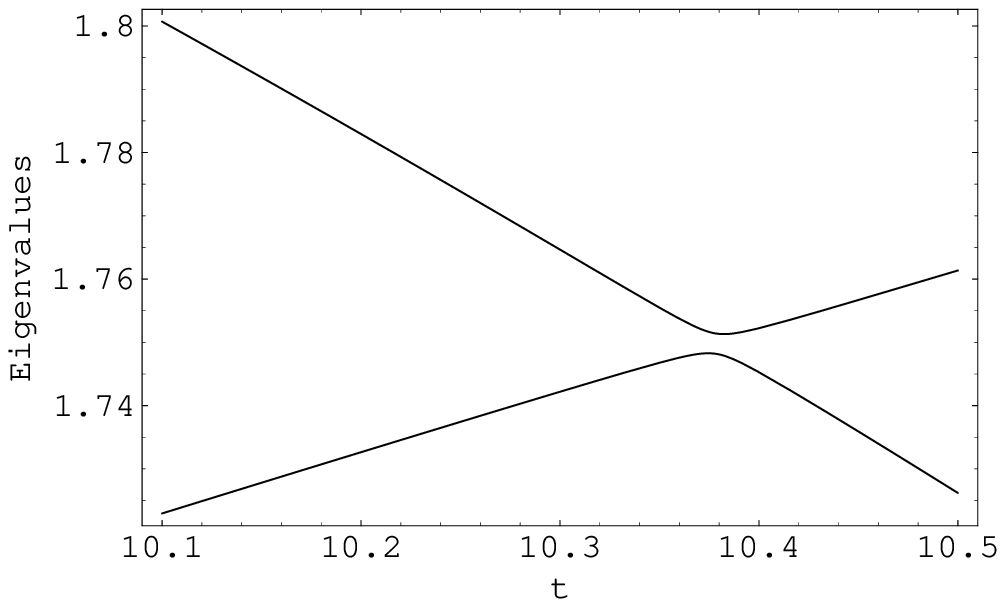}
\caption{\label{Fig4}An avoided crossing for the lowest two energy
levels for $\alpha =1$.	 Note the large variations in the first
derivatives of the branches.}
\end{center}
\end{figure}

A closer inspection of spectral flow (the flow of eigenvalues of
$\h(t)$ as functions of time) reveals a singular behaviour of the
flow also around $t\approx 10$, namely, that of an avoided crossing
of the instantaneous ground and first excited state as shown in
Figs.~\ref{Fig3} and~\ref{Fig4}.

\begin{figure}
\begin{center}
\includegraphics{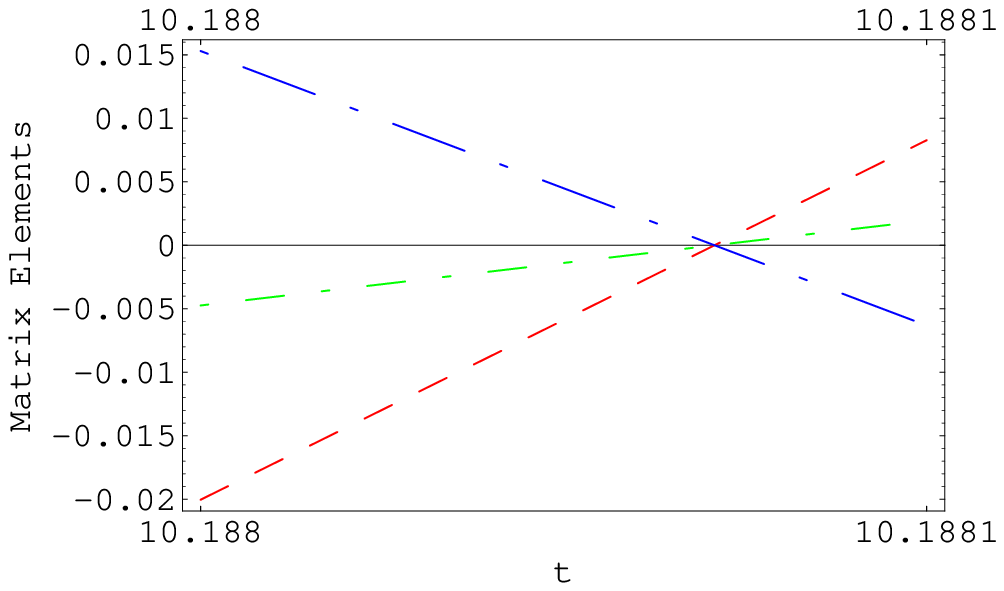}
\caption{\label{Fig5}The sudden change in occupation probabilities
in Fig.~\ref{Fig2} at certain time $t_0$ is due to the simultaneous
vanishing, at exactly the same time (and a bit earlier than $t_0$),
of the three different matrix elements
in~(\ref{condition}),~(\ref{condition1}) and~(\ref{condition2}) --
respectively corresponding to the (red) dash line, the (green)
dash-dot line, and the (blue) long dash-dot line.}
\end{center}
\end{figure}

As a matter of fact, all these anomalous behaviours are caused by
the violation of the condition~(\ref{condition}), or equivalently,
by the realisation of~(\ref{condition1}) and~(\ref{condition2}), as
shown by the simultaneous vanishing of the three matrix elements,
see Fig.~\ref{Fig5}, around (but a bit earlier than) the time of the
anomalous transfer of probability from the ground state to the first
excited state.

\subsection*{Appropriately large values for $\alpha$}
\begin{figure}
\begin{center}
\includegraphics{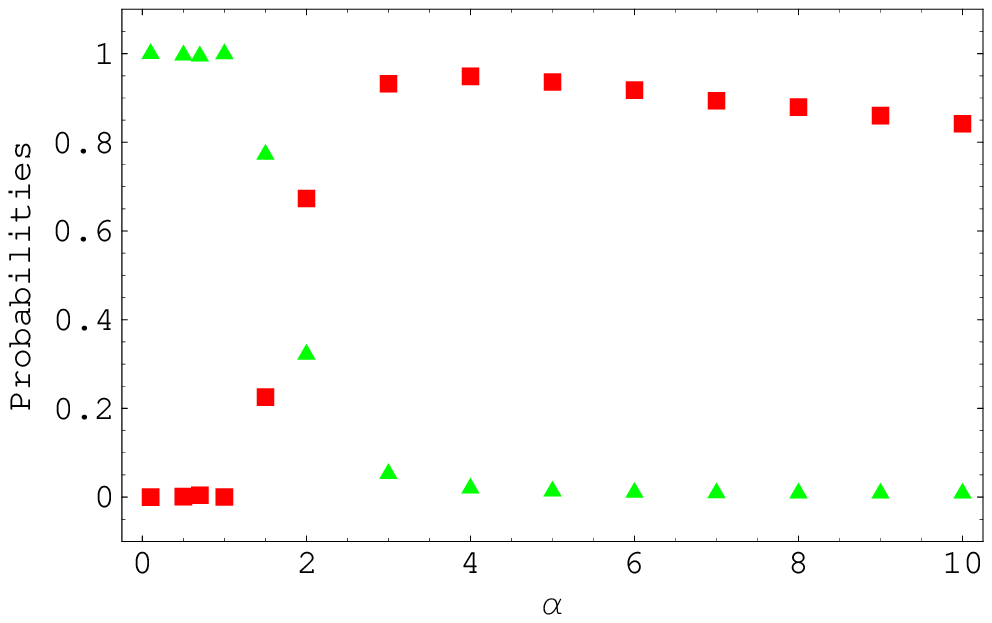}
\caption{\label{Fig6}The final occupation probabilities for the
ground state (red boxes) and first excited state (green triangles)
at $T=13.3444$ for different values of $\alpha$'s.  As expected, the
true ground state dominates for larger $\alpha$.}
\end{center}
\end{figure}

\begin{figure}
\begin{center}
\includegraphics{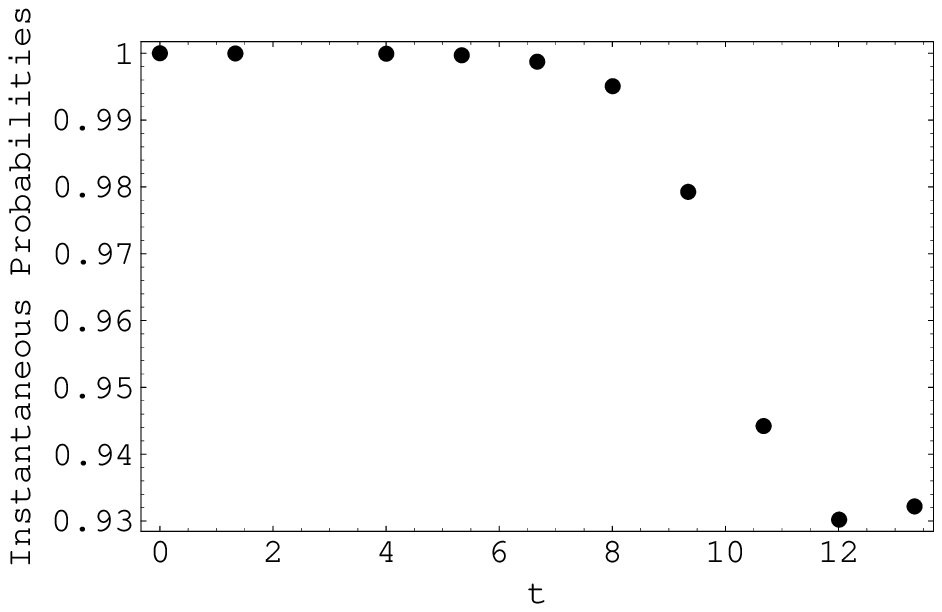}
\caption{\label{Fig7}The occupation probability of the instantaneous
ground state as a function of time, for $T=13.3444$ but with $\alpha
=3$.	(Note that the vertical scale does not go down to zero.)}
\end{center}
\end{figure}

\begin{figure}
\begin{center}
\includegraphics{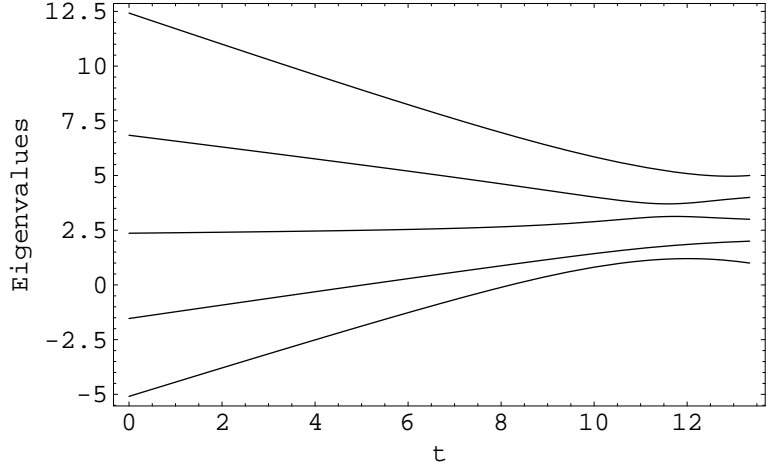}
\caption{\label{Fig8}The spectral flow shows no anomalous behaviour
for $\alpha = 3$.}
\end{center}
\end{figure}

\begin{figure}
\begin{center}
\includegraphics{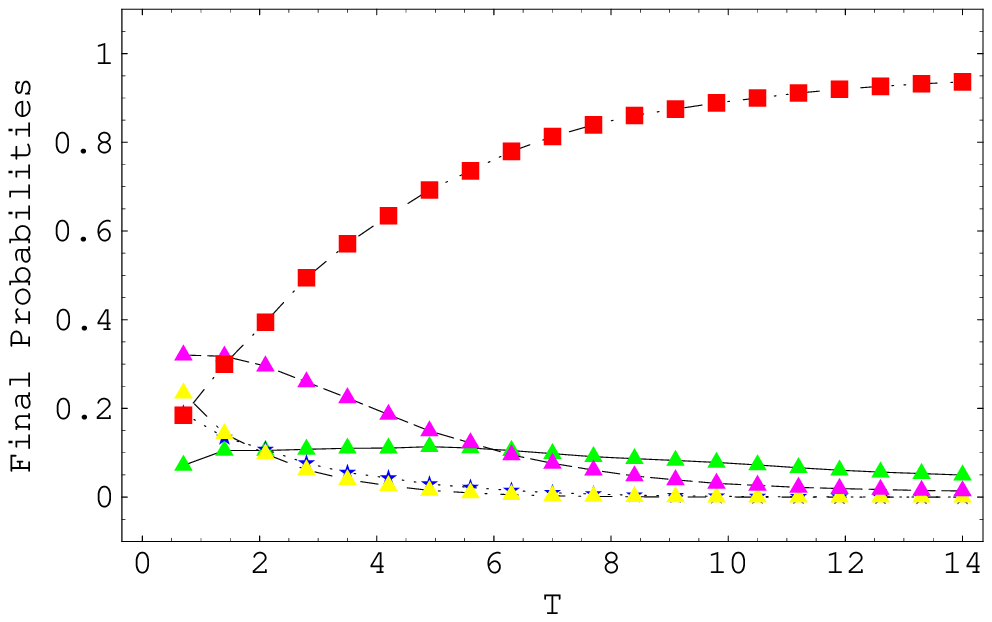}
\caption{\label{Fig9}The probabilities, for $\alpha = 3$, for
different Fock states as functions of the {\em final} time $T$.
There is only one state that has a probability eventually rising
above 0.5 (red boxes) and that is the true ground state of $H_P^S$,
namely $|4\rangle$.}
\end{center}
\end{figure}

Surely, for that value of $\alpha =1$ one could increase $T$ and
eventually will have an arbitrarily high occupation probability of
the final ground state, as guaranteed by the quantum adiabatic
theorem. However, for the present value of $T=13.3444$ we can avoid
the violation of~(\ref{condition}), and thus restore the
criterion~(\ref{criterion}), by widening the gap between the
instantaneous ground and first excited states.	This can be achieved
by choosing an appropriately large value for the parameter $\alpha$.
In Fig.~\ref{Fig6} we plot the occupation probabilities of the
instantaneous ground and first excited states as functions of
$\alpha$.  It is seen that as long as $\alpha > 2$ our
identification criterion~(\ref{criterion}) is again valid.
We present further results for $\alpha= 3$, as an example, in
Figs.~\ref{Fig7},~\ref{Fig8} and~\ref{Fig9}.  None of those shows
any anomalous behaviour.  (We also plot the ground state occupation
for $\alpha=3$ (filled black circles) in Fig.~\ref{Fig2} for
comparison with the case $\alpha=1$.)

We have provided some arguments in~\cite{thisissue} to indicate that
the condition~(\ref{condition}) can always be obtained for
sufficiently large values of $|\alpha|$. The larger this value is,
the more dominant role $H_I$ has in $\h(t)$, such that the lowest
branch of the spectral flow (that is, the branch of the ground
state) would be straightened out as the case in Fig.~\ref{Fig8}, and
any avoided crossing as the one in Figs.~\ref{Fig3} and~\ref{Fig4}
could not be formed.

How could one estimate such a value for $|\alpha|$?  One possible procedure is:\\
$\bullet$ We firstly choose a value for $\alpha$ and use the
criterion~(\ref{criterion}) to obtain a candidate state, say
$|m\rangle$.\\
$\bullet$ Then the diagonal term $\langle m|\h(t)|m\rangle$ is of
the form
\begin{eqnarray}
(1-t/T)(m +|\alpha|^2) + (t/T)\langle m|H_P|m\rangle.
\label{diagonal}
\end{eqnarray}
If we choose $\alpha$ such that the first term in~(\ref{diagonal})
dominates over the second term, for some $t$ close to $T$, then
$H_I$ would dominates over $H_P$ for this state and at this time. As
a result, no avoided crossing (which is due to the influence of
$H_P$) could be formed at this late $t$, and thus could never be
formed. A repeat of the adiabatic process and another application of
the criterion~(\ref{criterion}) would then reveal if the candidate
$|m\rangle$ is the true ground state. Otherwise, a different
candidate would be obtained and we could repeat the whole procedure
of choosing $\alpha$ again.

\subsection*{Sensitivity with different boundary conditions for the truncation}
In general, we suspect that the possibility of the violation
of~(\ref{condition}) is {\em an artefact of the truncation of the
Fock space} since it is extremely sensitive to the truncation and
its associated boundary conditions.	 Note also that any truncation
would distort the spacings between the eigenvalues of $H_I$,
severely away from unity particularly near the truncation.	In
Smith's truncation to five dimensions, for example, the four
spacings between the eigenvalues of $H_I^S$ are 1.11927, 1.40986,
1.84784, and 2.54043, respectively from the lowest to the highest
eigenvalues, which are $(0.0114457, 1.13072, 2.54058, 4.38841,
6.92885)$.	That is, the spacing between the fourth and fifth
eigenvalues near the truncation is 2.54043, which should have been
one. This and the fact that the ground state of $H_P^S$ happens to
be right at the truncation, $|n\rangle = |5\rangle$, make the
problem very sensitive to boundary effects.

However, the addition of extra states or the imposition of
(anti)-periodic boundary conditions on the truncated spaces may
remove the violation of~(\ref{condition}), as shown below.

\begin{figure}
\begin{center}
\includegraphics{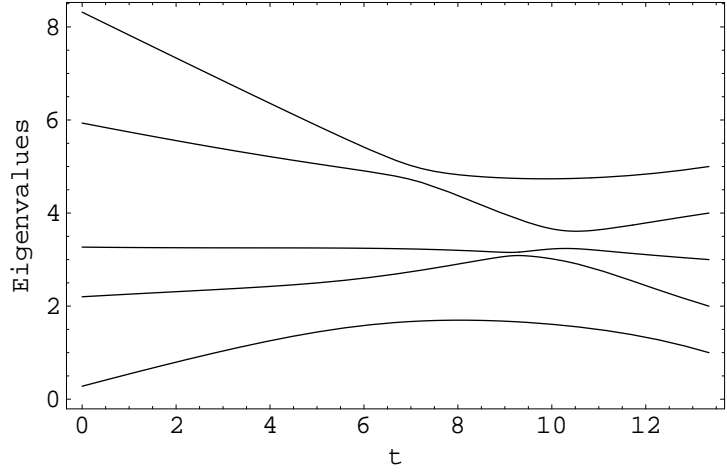}
\caption{\label{Fig10}The spectral flow for $\alpha = 1$, with
periodic condition for the truncated Fock space.}
\end{center}
\end{figure}
\begin{figure}
\begin{center}
\includegraphics{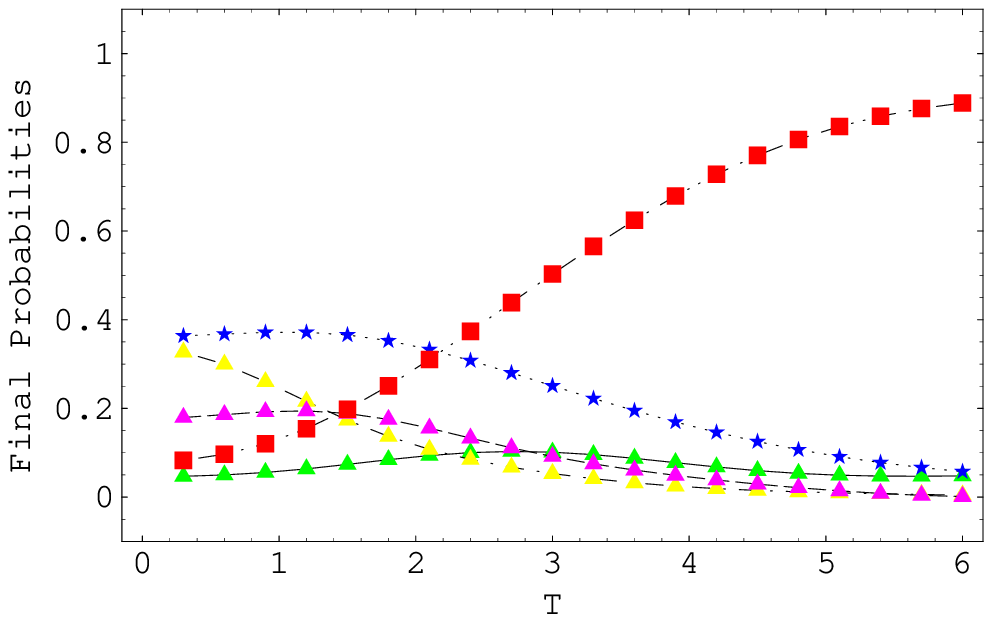}
\caption{\label{Fig11}The final occupation probabilities for
different Fock states at the final time $T$ for $\alpha = 1$ but
with periodic condition for the truncated Fock space. The boxes
represent the true final ground state $|4\rangle$.}
\end{center}
\end{figure}
\begin{figure}
\begin{center}
\includegraphics{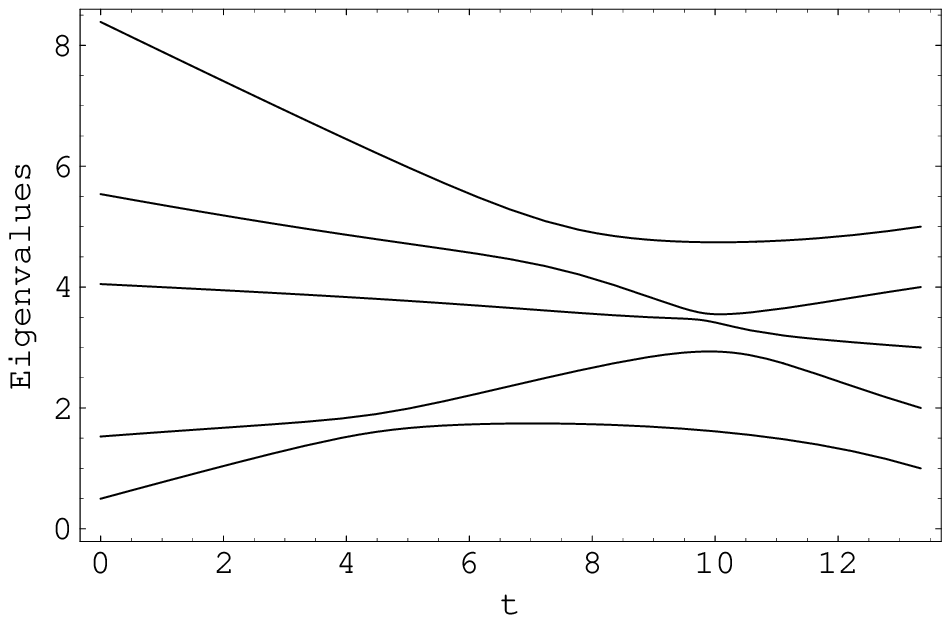}
\caption{\label{Fig12}The spectral flow for $\alpha = 1$, with
anti-periodic condition for the truncated Fock space.}
\end{center}
\end{figure}
\begin{figure}
\begin{center}
\includegraphics{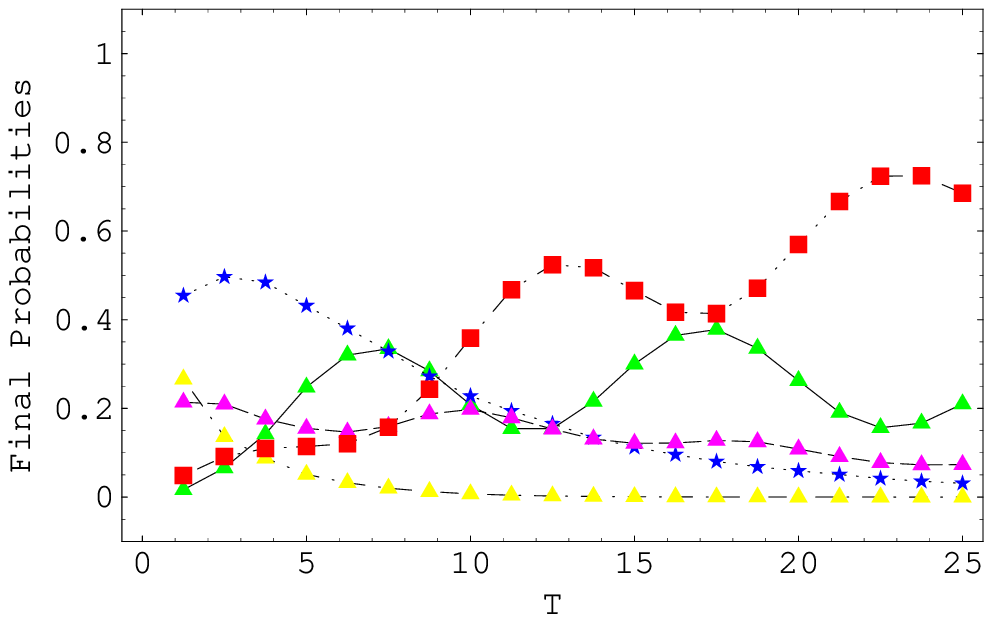}
\caption{\label{Fig12b}The final occupation probabilities for
different Fock states at the final time $T$ for $\alpha = 1$ but
with anti-periodic condition for the truncated Fock space.	The
boxes represent the true final ground state $|4\rangle$.}
\end{center}
\end{figure}

In obtaining the truncated matrices~(\ref{matrices}), Smith has
implicitly imposed the boundary condition
\begin{eqnarray}
a^\dagger \;|4\rangle = 0; && a |0\rangle = 0,
\end{eqnarray}
on the underlying truncated Fock space.  However, other boundary
conditions can also be imposed on an equal footing, for example,
\begin{eqnarray}
a^\dagger \;|4\rangle = \pm\sqrt{5}\;|0\rangle; && a \;|0\rangle =
\pm\sqrt{5}\;|4\rangle, \label{bcond}
\end{eqnarray}
where the plus (minus) sign corresponds to periodic (anti-periodic)
boundary condition on a Fock space having only five vectors,
$\{|0\rangle, |1\rangle, |2\rangle, |3\rangle, |4\rangle\}$. With
these new (anti)-periodic boundary conditions, we would have
\begin{eqnarray}
H_I^S &\to& \left(
\begin{array}{ccccc}
6& -\alpha^*& 0& 0& \pm\alpha^*\sqrt{5}\\
-\alpha& 2& -\alpha^*\sqrt{2}& 0& 0\\
0& -\alpha\sqrt{2}& 3& -\alpha^*\sqrt{3}& 0\\
0& 0& -\alpha\sqrt{3}& 4& -2\alpha^*\\
\pm\alpha\sqrt{5}& 0& 0& -2\alpha& 5
\end{array}\right);
\end{eqnarray}
with $a^\dagger a |0\rangle = 5 |0\rangle$, while assuming that
$H_P^S$ maintains the same form.
Figs.~\ref{Fig10},~\ref{Fig11},~\ref{Fig12} and~\ref{Fig12b} show
the non-anomalous and drastically different behaviour, as compared
to~(\ref{matrices}), for the same parameters $\alpha=1$ and
$T=13.3444$.

Based on such sensitivity and some other evidence, we suspect that a
violation of~(\ref{condition}) or, equivalently, a realisation
of~(\ref{condition1}) and~(\ref{condition2}) could not be attained
either for very large or an infinite number of dimensions, or for
more than one variable.	 It is perhaps appropriate to mention here
that all the numerical simulations presented in~\cite{thisissue}
have been obtained not with truncations with rigid boundary as the
cases Smith has considered but with movable boundaries where more
Fock states are included when and if necessary.	 Results from
further analytic investigations for an infinite number of dimensions
would be reported elsewhere when they become available\footnote{They
are now available in~\cite{Kieu2006} with a result confirming that,
indeed, the condition~(\ref{condition}) can be violated {\em
neither} in infinite dimensions, {\em nor} in a finite number of
dimensions with suitable boundary conditions, such as the
(anti)-periodic boundary conditions~(\ref{bcond}).}.

We should note further that~(\ref{criterion}) may not be the only
means which can be used to identify the final ground state. A more
general one, as discussed in~\cite{kieu-contphys, kieu-intjtheo},
could be a matching of the statistics of the measurement outcomes of
a physical quantum adiabatic process (in a space of presumably
infinite dimensions) with those of a numerical simulation of the
process with a necessarily truncated space.  As the size of the
truncation is enlarged, the true final ground state would eventually
be included in such a truncated Fock space. From then on, further
enlargement of the truncated Fock space only serves to improve the
precision of the statistics comparison. (This is another
manifestation of the finitely refutable nature of our problem, as
mentioned in~\cite{thisissue}.)

Finally, we also would like to point out that in addition to the
physical process discussed here, which is governed by {\em the
linear Schr\"odinger equation}, we can also map Hilbert's tenth
problem into a class of {non-linear differential
equations}~\cite{kieu-royal, Kieu2005}, that is readily available to
pure mathematical investigation.

\section*{Acknowledgments}
We are indebted to Warren Smith for numerous email exchanges and for
his critical observations that led to the counterexample, which in
turn has led to the investigation above.	Other criticisms by Smith
in~\cite{Smith2005} also overlap with others' elsewhere, to which we
have replied in~\cite{thisissue}.  This work has been supported by
the Swinburne University Strategic Initiatives.

\bibliography{c:/1data_16Apr05/papers/Computability}
\bibliographystyle{plain}

\end{document}